\documentclass[preprintnumbers,amsmath,11pt,amssymb,floatfix,superscriptaddress,nofootinbib]{article}

\def\mytitle#1{\setcounter{equation}{0}
\setcounter{footnote}{0}
\begin{flushleft}\Large\textbf{#1}\end{flushleft}
\vspace{0.25cm}}
\def\myname#1{\leftline{{\large #1}}\vspace{-0.13cm}}
\def\myplace#1#2{\small\begin{flushleft}\textit{#1}\\
\texttt{#2}\end{flushleft}}

\def\myclassification#1{\small\noindent
Keywords :
       #1\vspace{0.5cm}}
\usepackage{graphicx}% Include figure files
\begin{document}

\mytitle{Entropy for the Interior of a Schwarzschild Black Hole Assuming the Mass is Increasing With Time}

\myname{$Sandip~Dutta$\footnote{Duttasandip.mathematics@gmail.com } $and~Ritabrata~
Biswas$\footnote{biswas.ritabrata@gmail.com}}
\myplace{Department of Mathematics, The University of Burdwan, Golapbag Academic Complex, , City : Burdwan-713104, Dist. : Purba Burdwan, State : West Bengal, India}{} 
 
%%%%%%%%%%%%%%%%%%%%%%%%%%%%%%%%%%%%%%%%%%%%%%%%%%%%%%%%%%%%%%%%%%%%%%%%%%%%%%%%%%%%%%%%%%%%%%%%%%%%%%%%%%%%%%%%%%%%%%
\begin{abstract}
Black hole thermodynamics is the area of study that seeks to reconcile the laws of thermodynamics with the existence of black hole event horizons. Here we calculate the entropy corresponding to the interior of a Schwarzschild black hole for massless modes, assuming the mass of the black hole increasing with time. We find that the entropy is proportional to the Bekenstein-Hawking expression. Also, we can see that the evaluated entropy satisfies the second law of thermodynamics. Using the thermodynamic law we get a relation between temperature and inverse temperature. The special relativistic corrections to thermodynamic quantities are considered. The change in thermodynamic properties are analyzed when the velocity of the considered system is comparable to the speed of light. The effect of presence of scalar charge is incorporated.  \end{abstract}
%%%%%%%%%%%%%%%%%%%%%%%%%%%%%%%%%%%%%%%%%%%%%%%%%%%%%%%%%%%%%%%%%%%%%%%%%%%%%%%%%%%%%%%%%%%%%%%%%%%%%%%%%%%%%%%
\myclassification{Black hole physics; Thermodynamics; Thermodynamic processes; Entropy; Dynamic Black Hole}

%\newpage
%%%%%%%%%%%%%%%%%%%%%%%%%%%%%%%%%%%%%%%%%%%%%%%%%%%%%%%%%%%%%%%%%%%%%%
\section{Introduction}
There exist many theories regarding the compact objects like Black holes. Even the present day gravitational wave detections\cite{Abbott1} of black hole mergers can be treated as strong evidence of their existence. For the first time, theoretically Hawking has shown that black holes can evaporate and leave thermal radiation\cite{Hawking1, Hawking2}. This concept attracts people and different aspects of black hole thermodynamics\cite{Bardeen1} has been discussed. Many thermodynamic quantities were calculated. The study was analogous to the previous studies of classical thermodynamics. Entropy for a black hole was treated to be a quantity which is proportional to its ever increasing area of event horizon. Bekenstein-Hawking's entropy form\cite{Bekenstein1} is usually expressed as $$S_{BH}=\frac{A}{4 \hbar G},$$  where $A$ is the area of the black hole's event horizon.
The second law of thermodynamics states that the entropy increases with time. Another natural notion of entropy is the Von Neumann entropy given by $$S[\rho]=-tr(\rho~ln\rho),$$
where $\rho$ is the density matrix of a quantum system.
In classical level, we know that black hole mechanics follows the laws similar to the ordinary laws of thermodynamics\cite{Wald1}. A well known formula for the entropy of black hole in terms of Noether charge was later executed in the 90's\cite{Wald2}. Yet, a question has arrived as unanswered in the area of classical thermodynamics of black holes: what is the source of the entropy at the classical level? More generally, what are the classical microstates that corresponds to the entropy microstates? So, we found a long-standing issue in literature by this question, which has recently renewed: do extremal black holes have zero or non-zero entropy\cite{Carroll1}?
Recently, a new path of defining the inside volume of the Schwarzschild black hole has been introduced\cite{Christodoulou1} by Christodoulou and Rovelli (known as CR volume). This calculation was done for Kerr\cite{Bengtsson1}.
Now introducing the following inputs which taken in the paper\cite{Zhang1} without very clear description: (a) In expanding the Klein-Gordon equation under the time dependent background, the scalar field was chosen as $$exp\{-iET\}exp\{iI(\lambda,\theta,\phi)\},$$ where E is identified as the energy. (b) From the free energy we obtain the entropy, the derivative with respect to the inverse temperature was taken. (c) Finally, we use the flux relation(Hawking expression)\cite{Zhang1}.

Here we tried to get a clean knowledge about the above inputs. Express the interior volume as integral form\cite{Zhang1} and defined an effective metric. Then we evalute the Hamiltonian for a particle moving in this background. Finally, we use the Gibb's free energy for a massless particle and find the entropy as time dependent which is a very small impact on entropy changes.

The structure of the paper is as follows: firstly, we write a brief review of the results, given in \cite{Zhang1}. In section 3, we calculate the energy for a massless particle within the interior of the black hole. In section 4, we calculate the energy and in section 5, we consider a system with speed comparable to that of light and study the dependence of the corresponding thermodynamics as a function of the ratio $\frac{v}{c}$. Section 6 will study the effect of scalar charge present and finally we conclude in section 7.
%%%%%%%%%%%%%%%%%%%%%%%%%%%%%%%%%%%%%%%%%%%%%%%%%%%%%%%%%%%%%%%%%%%%%%%
\section{Review of Previous Works}
%%%%%%%%%%%%%%%%%%%%%%%%%%%%%%%%%%%%%%%%%%%%%%%%%%%%%%%%%%%%%%%%%%%%%%%%%%%
First we write the well known Schwarzschild metric in Eddington-Finkelstein coordinates
\begin{equation}\label{sch_Eddington_Finkelstein_coord}
ds^2=-fdv^2+2dvdr+r^2d\Omega^2~~,
\end{equation}
where the function $f=f(r,~t)=1-\frac{2M(1+a_1 t+a_2 t^2+ ... +a_n t^n)}{r}$ (where, $0 < a_i<< 1)$ and $v$ is advanced time defined as $v=t+\int\frac{dr}{f(r,t)}=t+r+2M(1+a_1 t+a_2 t^2+ ... +a_n t^n)\ln\left|r-2M(1+a_1 t+a_2 t^2+ ... +a_n t^n)\right|$. The units are taken as : $G=c=\hbar=\kappa_{B}=1$. Here $a_i$ are parameters which are very small such that $(1+a_1 t+a_2 t^2+ ... +a_n t^n)$ is near to $1$. Some important facts we have to note here : we considered the mass of the BH is increasing with time because matters get accreted towards the BH with time. Also, the universe's average density is very small. So, the BH mass increase very slowly. Because of this we consider that the partial derivative of $f(r,t)$ with respect to $t$ is equivalent to zero i.e., $\frac{\partial f}{\partial t}\approx 0$ by neglecting the small terms and so, Einstein's equations are satisfied. Also, the radial coordinates increment is very minute. So, we assume that the mass of the BH is $M(1+a_1 t+a_2 t^2+ ... +a_n t^n)$ after a time $t$, although mass of the BH is $m$ (at $t=0$). Also, we have considered that $r$ varies from $2M(1+a_1 t+a_2 t^2+ ... +a_n t^n)$ to zero. 
We can find some literatures as \cite{Lange11} where perturbation arising from Lindblad operators which describes the coupling to Markovian bath have been analysed. We can consider the time dependent generalised Gibb's ensembles to justify our modification of the mass. The concerned system can be treated as an open quantum system.

Besides, we are able to find the studies of time dependent black hole solutions in different referewnces. The reference \cite{Bak22}  has analysed a time dependent black hole solution utilizing AdS/CFT correspondence. Time dependent Hamiltonian approach can be seen in the reference  \cite{Majhi33}. 

As we have chosen strongly the variables $a_1$, $a_2$, .... , $a_n$to be positive, we see the mass to increase only. This will support the second law of thermodynamics very particularly \footnote{choosing these variables negative such that the second law of thermodybamics is violated is nothing but to chose an unphysical system.}.

Now, considering an transformation $v\rightarrow v(T, \lambda)$ and $r\rightarrow r(T, \lambda)$, we have
$$
ds^2=\left\{-f\left(\frac{\partial v}{\partial T}\right)^2+2\frac{\partial v}{\partial T}\frac{\partial r}{\partial T}\right\}dT^2+\left\{-f\left(\frac{\partial v}{\partial \lambda}\right)^2+2\frac{\partial v}{\partial \lambda}\frac{\partial r}{\partial \lambda}\right\}d\lambda^2+r^2d\Omega^2~~,
$$
considering the cross term as zero by taking the transformation properly. Assuming that the condition $-f\left(\frac{\partial v}{\partial T}\right)^2+2\frac{\partial v}{\partial T}\frac{\partial r}{\partial T}=-1$ is enforced, if the spherically symmetric hypersurface is considered as the direct product of a 2-sphere and an arbitrary curve parametrized by $\lambda$ in $v-r$ plane then we are able to find the hypersurface $\Sigma : T=constant$ \cite{Zhang1, Christodoulou1} where we have
$$ds^2_{\Sigma}=-dT^2+\left\{-f\dot{v}^2 +2\dot{v}\dot{r}\right\}d\lambda^2+r^2d\Omega^2~~.$$
The interior volume within the horizon can be written by the surface $\Sigma\equiv \gamma\times S^2$ on which metric can be written as
$$ds^2=\left\{-f\dot{v}^2 +2\dot{v}\dot{r}\right\}d\lambda^2+r^2d\Omega^2~~,$$
where, $r=r(\lambda)$ and $v=v(\lambda)$, $\lambda$ being an arbitrary parameter\cite{Christodoulou1}.

The volume can be written as
\begin{equation}
 V_{\Sigma} = 4\pi \int d\lambda \sqrt{r^4 \left\{-f\dot{v}^2 +2\dot{v}\dot{r}\right\}}~~.
\end{equation}
Now, the metric takes the form of an integrand of Lagrangian,
\begin{equation}
dS^2_{eff} =r^4 \left\{-f\dot{v}^2 +2\dot{v}\dot{r}\right\}d\lambda^2
\end{equation}
For this Lagrangian, the coordinates are $(r,~v)$ and the momenta are $(P_r, P_v)$ respectively. Now, $\int dr dv dP_rdP_v$ represent the phase space volume. In this paper we evaluate the entropy inside this phase space volume. In the process of evaluation of entropy in a quantum statistical way, firstly we have to calculate the Hamiltonian of a particle, restricted inside the volume.
\section{Calculation of the Hamiltonian of a Particle}
%%%%%%%%%%%%%%%%%%%%%%%%%%%%%%%%%%%%%%%%%%%%%%%%%%%%%%%%%%%%%%%%%%%%%%
Considering $m$ as the mass of a particle which moves in a space time with a back ground metric given by
\begin{equation}\label{Dynamic_BH_Ansatz}
ds_{ansatz}^{2}=g_{ab}dx^adx^b=-dt^2+r^4\left(-f(r,t)dv^2+2dvdr\right)~~,
\end{equation}
the action (which supposed to have the reparametrization symmetry) seems to be
\begin{equation}\label{Action1}
S=m\int_1^2 dS_{ansatz}=m\int_1^2\left(g_{ab}dx^adx^b\right)^{\frac{1}{2}}~~.
\end{equation}
The velocities of the particle is given by $u^a=\frac{dx^a}{d\tau}$, where, $\tau$ is an arbitrary parameter and $x^a=x^a(\tau)$. The path of the particle is then given by
\begin{equation}\label{Action2}
S=\int_1^2 {\cal L}d\tau=m\int_1^2\left(g_{ab}\frac{dx^a}{d\tau}\frac{dx^b}{d\tau}\right)^{\frac{1}{2}}d\tau~~.
\end{equation} 
Comparison of both sides clear states the Lagrangian to have the form ${\cal L}=m\left(g_{ab}\frac{dx^a}{d\tau}\frac{dx^b}{d\tau}\right)^{\frac{1}{2}}$.
By using the Euler Lagrange equation, we can easily find the equation of the motion of the system as
$$\frac{d^2x^a}{d\tau^2}+\Gamma^a_{bc}\frac{dx^b}{d\tau}\frac{dx^c}{d\tau}=0\Rightarrow \frac{du^a}{d\tau}+\Gamma^a_{bc}u^bu^c=0~~,$$
where $\Gamma^a_{bc}$ is the Christoffel symbol. Also the above geodesic equation is true for any space time. To evaluate the Hamiltonian which describes the whole dynamical system, firstly, we have to calculate the momenta of the system given by
\begin{equation}\label{9}
P_a=\frac{\partial {\cal L}}{\partial \dot{x}^a}=\frac{m^2}{\cal L}g_{ab}\frac{dx^b}{d\tau}~~.
\end{equation}
Therefore, the canonical Hamiltonian is 
\begin{equation}
H_c=P_a\frac{dx^a}{d\tau}-{\cal L}=0~~.
\end{equation}
In reparametrization invariant theory, canonical Hamiltonian was zero which is a typical signature of this theory; for Minkowski space time the same thing can be observed. For the present article, we have to analyse the Hamiltonian of the system  given by (\ref{Dynamic_BH_Ansatz}). For chronological gauge \cite{Hanson1}, in flat space time the analysis is done and for proper time gauge in \cite{Fulop1}. For calculation of entropy, we present the constraint which are independent.

Since the momenta are not independent, 
\begin{equation}
P^2=g^{ab}P_aP_b=m^2~~.
\end{equation}
Then we get primary constraint 
\begin{equation}\label{12}
\Phi=P^2-m^2\approx 0
\end{equation}
Using Dirac's algorithm \cite{Dirac1}, the primary constraint and the Hamiltonian are proportional to each other, therefore
\begin{equation}\label{14}
H_{T}=\xi (\tau) \Phi=\xi(\tau)\left(P^2-m^2\right)~~,
\end{equation}
where $\xi$ depends on $\tau$ and is known as proportionality constant.

Hence
\begin{equation}\label{15}
\dot{x}^a=\frac{dx^a}{d\tau}=u^a=\left\{x^a,~H_T\right\}=2\xi P^a
\end{equation}
and 
\begin{equation}
\dot{P}_a=\left\{P_a,~H_T\right\}=-\frac{\partial H}{\partial x^a}=-\xi \frac{\partial g^{bc}}{\partial x^a}P_bP_c
\end{equation}
also using (\ref{9}) and (\ref{15}) we get, $u^a=2\xi P^a=2\xi \frac{m^2}{{\cal L}}u^a$. So we have 
\begin{equation}\label{17}
\xi=\frac{{\cal L}}{2m^2}~~.
\end{equation}
Therefore, the total Hamiltonian is 
\begin{equation}
H_T=\frac{{\cal L}}{2m^2}\left(P^2-m^2\right)~~.
\end{equation}
So far, we have considered one constraint only (given by (\ref{12})). This characterises the system as of first class and hence it has gauge freedom \cite{Kawai1}. Imposing some conditions on the arbitrary parameter $\tau$ which we are going to interpret as proper time, the gauge freedom can be removed. A proper time gauge will be imposed throughout the subsequent analysis.

Therefore we are now with
\begin{equation}
\psi_2=\frac{P^0}{m}\tau-x^0\approx 0
\end{equation} 
and for primary constraint (\ref{14}) we have,
\begin{equation}
\psi_1=P^2-m^2\approx 0~~,
\end{equation} 
this makes the system as second class.

Therefore,
\begin{equation}\label{21}
\dot{\psi}_2=\frac{\partial \psi_2}{\partial t} +\left\{\psi_2,~H_T\right\}=0\Rightarrow\frac{P^0}{m}-\left\{x^0,~H_T\right\}+\frac{\tau}{m}\left\{P^0,~H_T\right\}=0~~.
\end{equation}
Also
\begin{equation}\label{22}
\left\{x^0,~H_T\right\}=\left\{x^0,~\xi P^2\right\}=2\xi g^{ab}\left\{x^0,~P_a\right\}P_b=2\xi P^0
\end{equation}
and
\begin{equation}
\{ P^0, H_T \}= \xi \left(2P^0 \frac{\partial g^{0b}}{\partial x^a} P_b - g^{0a}\frac{\partial g^{bc}}{\partial x^a}P_b P_c\right)~~.
\end{equation}
For the metric (\ref{Dynamic_BH_Ansatz}), we get 
$$\left\{P^0, H_T\right\}=0~~.$$
Therefore equation (\ref{21}) and (\ref{22}) gives 
\begin{equation}\label{24}
\xi=\frac{1}{2m}~~.
\end{equation}
From equations (\ref{17}) and (\ref{24}) we can obtain ${\cal L}=m$.

Also from equation (\ref{14}) and (\ref{15}) we get,
\begin{equation}
\dot{x}_a=\frac{1}{2m}\{x^a,P^2\}=\frac{P^a}{m}~~. 
\end{equation}
and 
\begin{equation}
\dot{P^a}=\frac{1}{2m}\{P^a,P^2 \} = \frac{1}{2m} \left(2\frac{\partial g^{ab}}{\partial x^c}g^{cd}-\frac{\partial g^{db}}{\partial x^c}g^{ca}\right)P_dP_b~~.
\end{equation}
\begin{equation}
\dot{P}^a=-\frac{1}{m}\Gamma^a_{bc}P^bP^c~~.
\end{equation}
here we useing the identity $$\partial_c g^{ab}=-g^{ad}g_{be}\partial_c g_{de}$$
for two dynamical variables, caculating the Dirac braket \cite{Bengtsson1, Majhi1} $$\{ f_1,f_2 \}^*$$ is 
\begin{equation}
\{ f_1,f_2 \}^*= \{ f_1,f_2 \} + \frac{1}{2P^0}\left(\{ f_1,P^2 \} \{ \psi_2,f_2 \}-\{ f_1,\psi_2 \} \{ P^2,f_2 \}\right)~~.
\end{equation}
for a dynamical variable, the equation of motion can be obtained by the following relation \cite{Dirac1} 
\begin{equation}
\dot{f_1} = \{ f_1, H \}^*~~.
\end{equation}
In fact $$H = P^0$$ serves our motive.
\begin{equation}
\dot{x}^a = \{ x^a, P^0 \}^* = g^{a0} + \frac{1}{P^0} \left(P^a + \frac{\tau}{m} g^{a0}  \Gamma_{bc}^0  P^b P^c \right)~~and~~
\dot{P}^a=\{P^a, P^0\}^*=\frac{\partial g^{ab}}{\partial x^c}g^{0c}P_b-\frac{1}{P^0}\Gamma^a_{bc}P^bP^c+g^{0a}\Gamma^0_{bc}P^bP^c
\end{equation}
By the gauge fixing constraint $$a=0$$ has been eliminated, then the equation of motion for the space component of $$a(a=\mu)$$  is 
\begin{equation}\label{32}
\dot{x}^\mu = \frac{P^\mu}{P^0}~~and~~\dot{P}^\mu = \frac{1}{P^0} \Gamma_{ab}^\mu P^a P^b
\end{equation}
From the above two equation, we can eliminate $P^0$ and get the desired geodesic equation. Now, 
\begin{equation}
g^{ab}P_aP_b=-\left(P^0\right)^2+2r^{-4}P_rP_v-fr^{-4}P_r^2=m^2
\end{equation}
The energy of a particle, i.e., the Hamiltonian is given by
\begin{equation}\label{34}
\epsilon=P^0=\left(-\frac{fP_r^2}{r^4}+\frac{2P_rP_v}{r^4}-m^2\right)^{\frac{1}{2}}
\end{equation}
We are calculating the energy for a massless particle (like photon). So we can consider $m\rightarrow 0$, then the equation (\ref{34}) can be written as
\begin{equation}\label{35}
\epsilon=\left(-\frac{fP_r^2}{r^4}+\frac{2P_rP_v}{r^4}\right)^{\frac{1}{2}}
\end{equation} 
\section{Entropy Calculation for Classical Gibb's free energy}
Now, we will construct the expression for entropy w.r.t. the energy for the massless particle. This entropy is defined inside the black hole. Also, no chemical potential terms are present, hence Gibb's free energy is $G_0=-\frac{1}{\beta}\ln Z=\frac{1}{\beta}\Sigma_\epsilon\ln \left(1-exp\{-\beta \epsilon\}\right)$, where $Z$ is the grand canonical partition function and $\beta$ is the inverse temperature. Gibb's free energy is not a relativistic invariant. The increment in mass can be considered as a slow process. So, we can consider that the effect of Gibb's free energy is negligible, i.e., the differences between Gibb's free energy before and after the change of hypersurface will be almost the same. In horizon, for maximum volume of the interior, we get the fixed value of the radial coordinate, i.e., $\dot{r}=0$, where $r$ can be obtained from the equation (\ref{32}) which implies $\dot{r}=r^{-4}\frac{\left(P_v+fP_r\right)}{P^0}=0\Rightarrow P_v+fP_r=0$. Again $r$ and ingoing null coordinates are depending on $\lambda$(a parameter)\cite{Christodoulou1}, i.e., $r=r(\lambda)$ and $v=v(\lambda),~~v=F(r)$, a function of $r$. Therefore the Gibb's free energy 
$$G_0=\frac{1}{\beta}\int dP_rdP_vdrdv \ln \left(1-exp\{-\beta \epsilon\}\right)\times \delta \left(P_v+fP_r\right)\delta (v-F(r))~~.$$
Using Dirac-Delta functions
$$G_0=\frac{1}{\beta}\int dP_rdr\ln\left[1-exp\left\{-\beta\left(-\frac{fP_r^2}{r^4}\right)^{\frac{1}{2}}\right\}\right]~~,$$
where $P_r$ varies from $0$ to $\infty$.

So
$$G_0=\frac{1}{\beta}\int dr\int_0^{\infty} \ln\left[1-exp\left\{-\beta\left(-\frac{fP_r^2}{r^4}\right)^{\frac{1}{2}}\right\}\right]$$
incorporating $\frac{\beta P_r \sqrt{-f}}{r^2}=x$ we have 
\begin{equation}\label{Gibbs equation1}
G_0=-\frac{\pi^2}{6\beta^2}\int \frac{r^2}{\sqrt{-f}}dr
\end{equation}
Let us consider as time goes the radius of the black hole increases, then the radial coordinate varies from $2M(1+a_1 t+a_2 t^2+ ... +a_n t^n)$ to zero. So, we have 
$$G_0=-\frac{\pi^2}{6 \beta^2}\int_{2M(1+a_1 t+a_2 t^2+ ... +a_n t^n)}^0\frac{r^2}{\sqrt{\frac{2M(1+a_1 t+a_2 t^2+ ... +a_n t^n)}{r}-1}}dr~~.$$
Incorporating $y=\frac{r}{2M(1+a_1 t+a_2 t^2+ ... +a_n t^n)}$, we have 
$$G_0=\frac{5\pi^3}{12 \beta^2}M^3(1+a_1 t+a_2 t^2+ ... +a_n t^n)^3$$

Hawking showed that quantum effects allow black holes to emit exact black body radiation\cite{Hawking1}. The electromagnetic radiation is produced as if emitted by a black body with a temperature inversely proportional to the mass of the black hole. Also, our system is like black body which encloses massless particles. Hence the particles are also at inverse temperature $\beta=8\pi M$.
Therefore, the Gibb's free energy
\begin{equation}\label{35a}
G_0=\frac{5}{6144} \beta(1+a_1 t+a_2 t^2+ ... +a_n t^n)^3
\end{equation}
Hence, the entropy is 
\begin{equation}\label{35b}
S_0=\beta^2\frac{\partial G_0}{\partial \beta}=\frac{5 \beta^2}{6144}(1+a_1 t+a_2 t^2+ ... +a_n t^n)^3
\end{equation}

So, we can say that the entropy is monotonically increasing as time increases but the increment of entropy depends on time which is very small.

In this connection we wish to reestablish the result given by \cite{Planck1}, where they have proved that Planck entropy is a Lorentz invariant. The empirical temperature in relativistic thermodynamics is a Lorentz invariant\cite{Przanowski1}. Let us assume in the proper frame, $T_0$, $\delta Q_0$ and $\delta L_0$ are the absolute temperature, the amount of heat entering the system and the thermodynamic work done over the system respectively. For reversible processes
\begin{equation}\label{36}
\delta Q_0 = T_0 dS_0.
\end{equation} 
At the beginning if we had assumed that the efficiency of the Carnot cycle is an invariant, then the entropy is an invariant, i.e., 
\begin{equation}\label{37}
dS = \frac{\delta Q}{T} = \frac{\delta Q_0}{T_0} = dS_0.
\end{equation}
By the second law of thermodynamics the Pfaffian form $\delta Q_0$ depending only on the empirical temperature. This leads to the Clausius equality
\begin{equation}\label{38}
\oint \frac{\delta Q_0}{T_0} = 0
\end{equation}
for every cyclic reversible process.

Let us assume that the second law of thermodynamics is in force in relativistic thermodynamics in any inertial frame. Hence, the absolute temperature of the form
\begin{equation}\label{39}
T=T(T_0,v)~~,~~\lim_{v \to 0} T(T_0,v) = T_0~~,
\end{equation}
where the thermodynamic system moving with a constant velocity $\vec{v}$ with respect to a frame $K$. Also, the Clausius equality in $K$ reads
\begin{equation}\label{40}
\oint \frac{\delta Q}{T} = 0.
\end{equation}
Again, the relativistic transformation of heat found by H. Ott\cite{Ott1} and independently by H. Arzelies\cite{Arzelies1} and C. M$\phi$ller\cite{Møller1, Møller2}
\begin{equation}\label{41}
\delta Q = \gamma(v) \delta Q_0
\end{equation}
(see also \cite{Møller3}).

Substituting (\ref{41}) and (\ref{39}) into (\ref{40}), we obtain,
\begin{equation}\label{42}
\oint \frac{\delta Q_0}{\{\gamma(v)\}^{-1} T(T_0,v)} = 0.
\end{equation}
Comparing (\ref{42}) and (\ref{38}), we get,
\begin{equation}\label{43}
\{\gamma(v)\}^{-1} T(T_0,v) = bT_0,
\end{equation}
where $b$ is an arbitrary constant. Taking $v \rightarrow 0$ in the equation (\ref{43}) and applying (\ref{39}), we get, $b=1$. So finally
\begin{equation}\label{44}
T= \gamma(v) T_0.
\end{equation}
This is the second law of relativistic thermodynamics for reversible processes and the relativistic transformation of absolute temperature consistent with (\ref{41}). This transformation rule and the transformation law are different\cite{Einstein1, Planck1, Mosengeil1, Pauli1, Laue1}
\begin{equation}\label{45}
T^{(Planck)}= \{\gamma(v)\}^{-1} T_0
\end{equation}
following from the transformation rule\cite{Landsberg1, Landsberg2, Landsberg3, Landsberg4}
\begin{equation}\label{46}
T^{(L)}= T_0
\end{equation}
consistent with 
\begin{equation}\label{47}
\delta Q^{(L)} = \delta Q_0.
\end{equation}
Now, from the equation (\ref{36}), (\ref{41}) and (\ref{44}), we get, 
\begin{equation}\label{48}
\delta Q = T dS_0 
\end{equation}
which implies that according to Planck\cite{Planck1} entropy is a Lorentz invariant
\begin{equation}
S=S_0
\end{equation}

Some important things we have to mention here : firstly, we have considered that the massless modes inside the horizon are at a temperature whose value is same as that of the event horizon. We should remember that $r$ varies from 0 to $2M$ but our consideration was slightly different than this. We have assumed the mass of the black hole is increasing as time increases because matters get accreted towards a black hole with time. But the average density of the universe is very small. So the increment of the mass of the black hole is very slow. For this we assumed that the radial coordinate's increment also is very minute. This leaded us to the consideration that the mass of the black hole is $M(1+a_1 t+a_2 t^2+ ... +a_n t^n)$ after a time $t$ while $M=M(t=t_0=0)$. We have also assummed that $r$ varies from $2M(1+a_1 t+a_2 t^2+ ... +a_n t^n)$ to zero.

Also, Hawking showed that black holes emit thermal Hawking radiation\cite{Bekenstein1, Hawking3} corresponding to a certain temperature(Hawking temperature)\cite{Hawking1, Hawking4}. Using the thermodynamic relationship between energy, temperature and entropy, Hawking was able to confirm Bekenstein's conjecture and fix the constant of proportionality at $1/4$\cite{Hawking5}:
 
$S_{BH}=\frac{A}{4 }$ is the entropy on the horizon of the black hole, where $A=16\pi M^2$ is the area of the event horizon.

Hence, the relation between $S_0$ and $S_{BH}$ is $$S_0=\frac{5\pi}{384}(1+a_1 t+a_2 t^2+ ... +a_n t^n)^3 S_{BH}$$

Thus we can see that the entropy on the horizon is greater than the entropy inside the black hole i.e. $S_{BH} > S_0$. Also the entropy inside the black hole is proportional to the time($t$), i.e., if we increase the time then the entropy($S_0$) inside the black hole increase.    

The first law of thermodynamics provides the basic definition of internal energy, associated with all thermodynamics systems, and states the rule of conservation of energy. The second law is concerned with the direction of natural processes. It asserts that a natural process runs only in one sense, and is not reversible. Also, we can see from the above $$\frac{dS_0}{dt}>0$$.

Hence, the total entropy can never decrease over time and the process is irreversible.

According to the first law of thermodynamics 
\begin{equation}\label{first1}
dM=T_0dS_0
\end{equation}
$$\Rightarrow \beta= \frac{768}{15 \pi T_0} \frac{1}{(1+a_1 t+a_2 t^2+ ... +a_n t^n)^2},$$

where $\beta$ is the reciprocal of the thermodynamic temperature of the system\cite{Castle1, Garrod1}. Thermodynamic beta is essentially the connection between the information of a physical system through its entropy and the thermodynamics associated with its energy.

{\bf Case:I}

If $a_i=0$ for $i>1$ then $(1+a_1 t+a_2 t^2+ ... +a_n t^n) \approx (1+a_1 t)$ and 

$S_0=\frac{5\pi}{384}(1+a_1 t)^3 S_{BH}$ and $\beta= \frac{768}{15 \pi T_0} \frac{1}{(1+a_1 t)^2}$

\begin{figure}[ht]
\begin{center}

~~~~~~~Fig.1a~~~~~~~~~~~~~~~~~~~~~\\
\includegraphics[height=2.5in, width=3.2in]{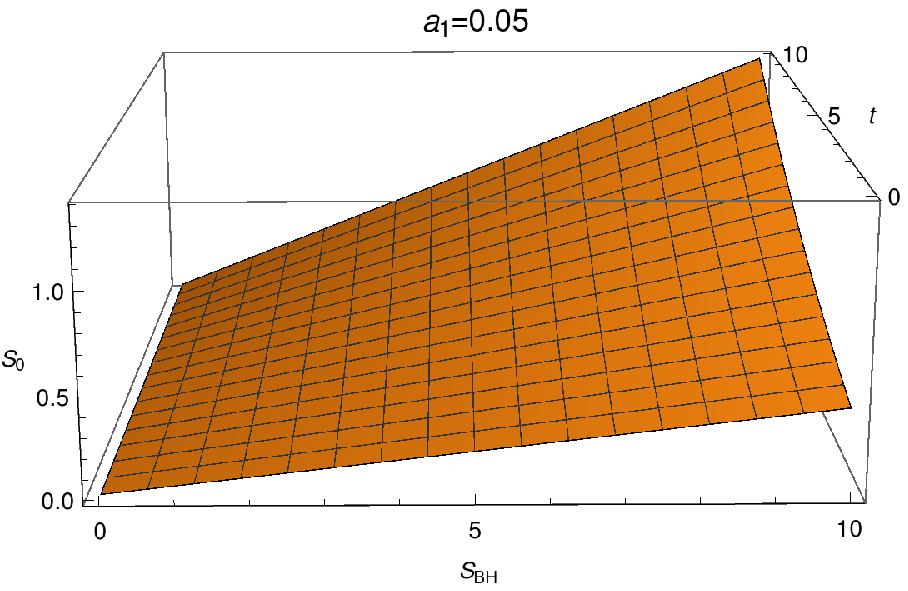}\\
\vspace{.1cm}
Relation between $S_0$, $S_{BH}$ and $t$

\end{center}
\end{figure}
If we increase the entropy on the horizon of the black hole ($S_{BH}$) then we show that the entropy inside the black hole increases here the time assumed as a constant. On the otherhand, if we variate the time then the rate of increasing of entropy inside the black hole become higher than the above case.
\begin{figure}[ht]
\begin{center}

~~~~~~~Fig.1b~~~~~~~~~~~~~~~~~~~~~\\
\includegraphics[height=2.5in, width=3.2in]{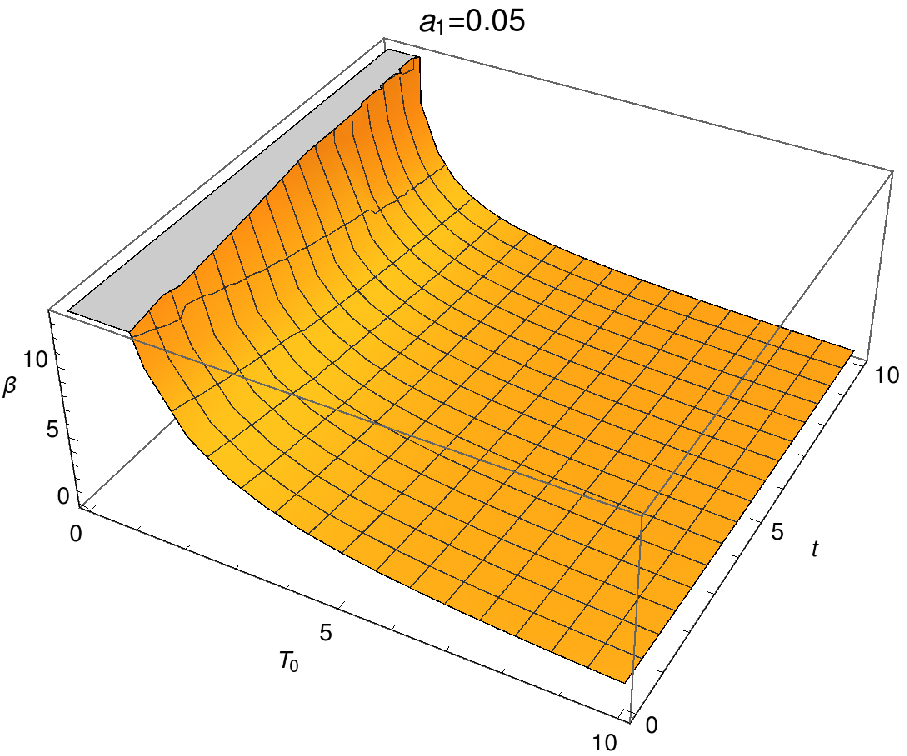}\\
\vspace{.1cm}
Relation between $\beta$, $T_0$ and $t$

\end{center}
\end{figure}

The relation between $\beta$ and $T_0$ is hyperbolic through out the figure, when time is not constant the eccentricity increase with time.
\newpage
{\bf Case:II}

If $a_i=0$ for $i>2$ then  $(1+a_1 t+a_2 t^2+ ... +a_n t^n) \approx (1+a_1 t+a_2 t^2)$ and

$S_0=\frac{5\pi}{384}(1+a_1 t+a_2 t^2)^3 S_{BH}$ and $\beta= \frac{768}{15 \pi T_0} \frac{1}{(1+a_1 t+a_2 t^2)^2}$
\begin{figure}[ht]
\begin{center}

~~~~~~~Fig.2a~~~~~~~~~~~~~~~~~~~~~\\
\includegraphics[height=2.5in, width=3.2in]{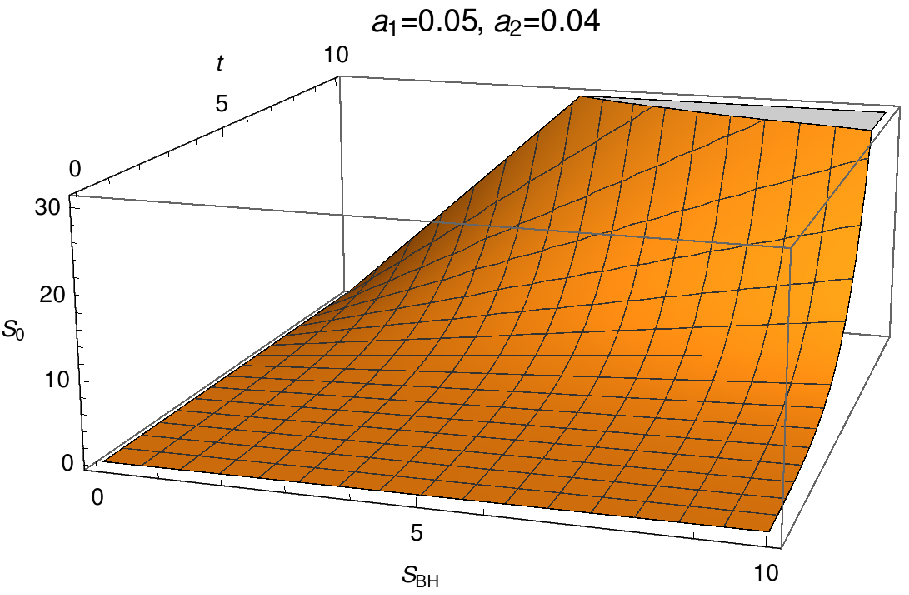}\\
\vspace{.1cm}
Relation between $S_0$, $S_{BH}$ and $t$

\end{center}
\end{figure}
When time constant the entropy inside the black hole increases with the increament with the horizon entropy. But if the time is not constant the rate of increasing entropy inside the black hole is distincity high.
\begin{figure}[ht]
\begin{center}

~~~~~~~Fig.2b~~~~~~~~~~~~~~~~~~~~~\\
\includegraphics[height=2.5in, width=3.2in]{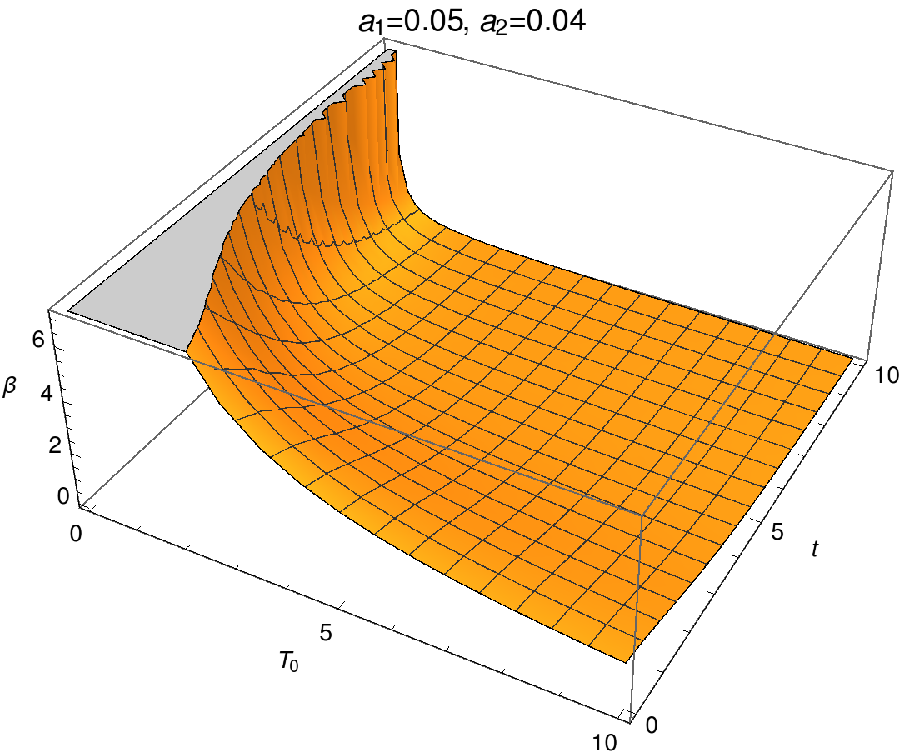}\\
\vspace{.1cm}
Relation between $\beta$, $T_0$ and $t$

\end{center}
\end{figure}
Here the rate of increasing of the eccentricity is higher than the Case:I.
\newpage
{\bf Case:III}

If $a_i=0$ for $i>1$ then  $(1+a_1 t+a_2 t^2+ ... +a_n t^n) \approx (1+a_1 t+a_2 t^2+a_3 t^3)$ and

$S_0=\frac{5\pi}{384}(1+a_1 t+a_2 t^2+a_3 t^3)^3 S_{BH}$ and $\beta= \frac{768}{15 \pi T_0} \frac{1}{(1+a_1 t+a_2 t^2+a_3 t^3)^2}$.
\begin{figure}[ht]
\begin{center}

~~~~~~~Fig.3a~~~~~~~~~~~~~~~~~~~~~\\
\includegraphics[height=2.5in, width=3.2in]{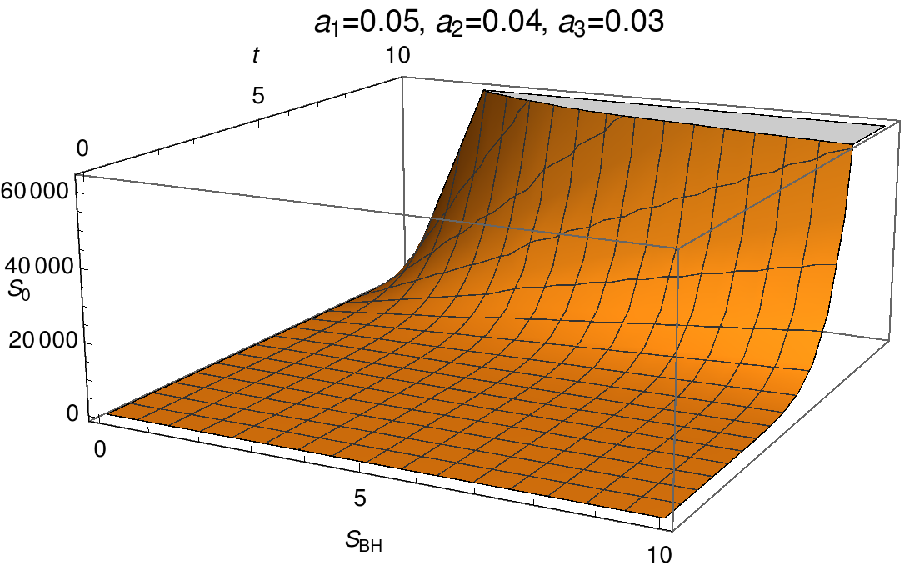}\\
\vspace{.1cm}
Relation between $S_0$, $S_{BH}$ and $t$

\end{center}
\end{figure}
If the time is not constant the rate of increasing entropy inside the black hole is distincity high from the Case:I and Case:II.
\newpage
\begin{figure}[ht]
\begin{center}

~~~~~~~Fig.3b~~~~~~~~~~~~~~~~~~~~~\\
\includegraphics[height=2.5in, width=3.2in]{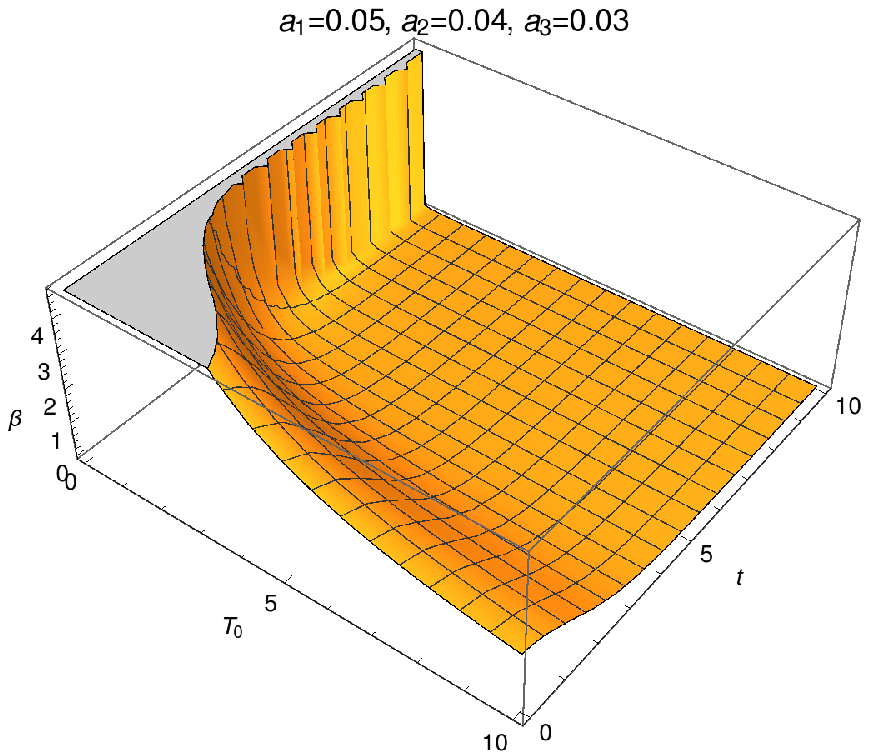}\\
\vspace{.1cm}
Relation between $\beta$, $T_0$ and $t$

\end{center}
\end{figure}
Here the rate of increasing of the eccentricity is higher than the Case:I and Case:II.

\section{Phenomenological Relativistic Thermodynamics}
%%%%%%%%%%%%%%%%%%%%%%%%%%%%%%%%%%%%%%%%%%%%%%%%%%%%%%%%%%%%
In proper inertial frame (i.e., say $K_0$), the relativistic counter part of famous Maxwell's distribution for ideal gas was proposed by F. Juttner\cite{Laue1}. Throughout a long period, Juttner distribution had been accepted by many literatures\cite{Pauli1, Pathria1, Pathria2, Synge1, Haar1, Groot1}. When a vessel, kept at labaratory, contains gas moving with constant velocity $\vec{v}$ with respect to the labaratory frame $K$, the similar assumptions and deductions like Juttner has been constructed. Some literature based on computer simulations\cite{Cubero1, Dunkel1, Rasinariu1, Montakhab1, Ghodrat1}, supports partially the correctness of the Juttner formula and its generalization. Some other distributions also have been proposed. Though Juttner distribution is found to be an efficient one\cite{Peano1}. The relativistic Gibb's distribution for an ideal gas contained in the vessel moving with a constant velocity $\vec{v}$ and 4-velocity $u^j=\left(\gamma(v)\frac{\vec{v}}{c}, \gamma(v)\right)$ turns to be 
\begin{equation}\label{ss1}
d \omega= \frac{1}{(2\pi )^{3N}N! Z}exp\{-\beta c u_j {\cal{P}}^j\}d^{3N}pd^{3N}q~~~,
\end{equation}
where $N$ is the number of particles, $\beta = \frac{1}{\kappa T_0}$ with $\kappa$ being the Boltzmann constant, ${\cal{P}}^j$ is the total momentum of the gas, $d^{3N}pd^{3N}q$ is the phase space volume element and $Z$ denotes the partition function
\begin{equation}\label{ss2}
Z= \frac{V^N}{(2\pi )^{3N}N!} \int_{R^{3N}} exp\{-\beta c u_j {\cal{P}}^j\}d^{3N}p~~~.
\end{equation} 
Again, the measure $d^3p$ transforms as follows\cite{Pathria2}
\begin{equation}\label{ss3}
d^3p= \gamma(v)\left(1 + \frac{\vec{v}\cdot \vec{p}_0}{c\sqrt{\vec{p}_0^2+m^2 c^2}} \right)d^3p_0~~~.
\end{equation}
Hence,
$$Z= \frac{V^N}{(2\pi )^{3N}N!} \left(\int_{R^{3N}} exp\{-\beta c u_j p^j\}d^{3}p\right)^N$$
$$=\frac{V^N}{(2\pi )^{3N}N!} \left(\int_{R^{3N}} exp\Big\{-\beta c \sqrt{\vec{p}_0^2+m^2 c^2} \Big\}\left[\gamma(v)\left(1 + \frac{\vec{v}\cdot \vec{p}_0}{c\sqrt{\vec{p}_0^2+m^2 c^2}} \right)\right] d^{3}p_0\right)^N$$
\begin{equation}\label{ss4}
=\frac{V_0^N}{(2\pi )^{3N}N!} \left(\int_{R^{3N}} exp\Big\{-\beta c \sqrt{\vec{p}_0^2+m^2 c^2} \Big\}d^{3}p_0\right)^N=Z_0 ,
\end{equation}
where the subindex `$0$' corresponds to the proper frame $K_0$. From the above equation, we can say that the partition function $Z$ is a Lorentz invariant\cite{Pathria2}.
 
Finally, the relativistic Gibbs distribution (\ref{ss1}) turns to be
\begin{equation}\label{ss5}
\omega_n=\frac{1}{Z}exp\{-\beta c u_j {\cal{P}}_n^j\}~~~~, ~~~n=1,~2,...
\end{equation}  
where the subindex `$n$' denotes a quantum state.
Also, from the first law of relativistic thermodynamics (for ideal liquid)\cite{Landau1, Przanowski1}
\begin{equation}\label{p1}
dE = TdS-pdV+\{\gamma(v)\}^2 \frac{v^2}{c^2} Vdp, 
\end{equation}
From (\ref{p1}), one can easily find
\begin{equation}\label{p2}
d\left[E- \{\gamma(v)\}^2 \frac{v^2}{c^2}V_p\right]=TdS-\{\gamma(v)\}^2 p dV
\end{equation}
Hence,
\begin{equation}\label{p3a}
T= \left(\frac{\partial E}{\partial S}\right)_V - \{\gamma(v)\}^2 \frac{v^2}{c^2} V\left(\frac{\partial p}{\partial S}\right)_V ,
\end{equation}
\begin{equation}\label{p3b}
p- \{\gamma(v)\}^2 \frac{v^2}{c^2} V \left(\frac{\partial p}{\partial V}\right)_S=-\left(\frac{\partial E}{\partial V}\right)_S
\end{equation}
and
\begin{equation}\label{p4}
\left(\frac{\partial T}{\partial V}\right)_S = -\{\gamma(v)\}^2 \left(\frac{\partial p}{\partial S}\right)_V
\end{equation}
Therefore, from the equation (\ref{p1}) and (\ref{p2}), we obtain,
\begin{equation}\label{p5}
T= \{\gamma(v)\}^2 \left(\frac{\partial E}{\partial S}\right)_{V,P}
\end{equation}
Also, pressure is a Lorentz invariant\cite{Przanowski1}, i.e., $p=p_0$ and as the volume  $V_0=\gamma(v) V$ we can write 
\begin{equation}\label{p6}
{\cal{H}} = E+pV= \gamma(v) \left(E_0+p_0 V_0\right) = \gamma(v) {\cal{H}}_0 ,
\end{equation}
where ${\cal{H}}$ is the enthalpy.

Using the definition of enthalpy and equation (\ref{p2}), we get,
\begin{equation}\label{p7}
T= \left(\frac{\partial H}{\partial S}\right)_p
\end{equation}
Therefore, the free energy $F$
\begin{equation}\label{p8}
F = E-TS- \{\gamma(v)\}^2 \frac{v^2}{c^2} pV = \gamma(v) F_0,
\end{equation}
where $F_0=E_0-T_0 S_0$ is the free energy of the system in its proper frame $K_0$.

Using the equation (\ref{p8}) into the equation (\ref{p2}), we get,
\begin{equation}\label{p9}
dF=-SdT-\{\gamma(v)\}^2 pdV
\end{equation}
Therefore, 
\begin{equation}\label{p10}
S=-\left(\frac{\partial F}{\partial T}\right)_V,~~\{\gamma(v)\}^2p=-\left(\frac{\partial F}{\partial V}\right)_T,~~E=F-T\left(\frac{\partial F}{\partial T}\right)_V - \frac{v^2}{c^2}V \left(\frac{\partial F}{\partial V}\right)_T
\end{equation}
and the relativistic Maxwell identities
\begin{equation}\label{p11}
\left(\frac{\partial S}{\partial V}\right)_T = \{\gamma(v)\}^2 \left(\frac{\partial p}{\partial T}\right)_T
\end{equation}
We define $G$ as the Gibbs function in a standard way 
\begin{equation}\label{p12}
G=E-TS+pV= \gamma(v) G_0
\end{equation} 
with $G_0=E_0-T_0 S_0+p_0 V_0$

Again, some invariant thermodynamical quantities are represented by the Massieu function\cite{Carrera1},
\begin{equation}\label{p13}
 \frac{G}{T}=\frac{G_0}{T_0}
\end{equation}
Using the equations (\ref{35a}) and (\ref{p12}), we get,
\begin{equation}\label{p14}
G= \frac{5\gamma(v)}{6144} \beta(1+a_1 t+a_2 t^2+ ... +a_n t^n)^3
\end{equation}
Therefore, the entropy is 
\begin{equation}\label{p15}
S=\beta^2 \frac{\partial G}{\partial \beta} =\frac{5\gamma(v) \beta^2}{6144}(1+a_1 t+a_2 t^2+ ... +a_n t^n)^3=\gamma(v)S_0
\end{equation}
According to the first law of thermodynamics and using the equation (\ref{p15}), we get,
\begin{equation}\label{p16}
dM=TdS=\gamma^{-1} T_0 \gamma dS_0=T_0dS_0,
\end{equation}
since $T=\gamma(v)^{-1} T_0$ . The above equation is same as the equation (\ref{first1}).

(see also \cite{Carrera1}).
 
$S_{BH}=\frac{A}{4}$ is the entropy\cite{Hawking5} on the horizon of the black hole, where $A=16\pi M^2$ is the area of the event horizon. Hence, the relation between $S$ and $S_{BH}$ is 
\begin{equation}\label{p17}
S=\frac{5\pi \gamma(v)}{384}(1+a_1 t+a_2 t^2+ ... +a_n t^n)^3 S_{BH}~~~,
\end{equation} 
where $\gamma(v)=\frac{1}{\sqrt{1-v^2/c^2}}$ and $v$ is the velocity between both the rest and moving reference frames. Let us assume $\frac{v}{c}=\alpha$, where $0\leq \alpha < 1$.
So, from the equation (\ref{p17})
\begin{equation}
S=\frac{5\pi \sqrt{1-\alpha^2}}{384}(1+a_1 t+a_2 t^2+ ... +a_n t^n)^3 S_{BH}
\end{equation}
{\bf Case:I}

If $a_i=0$ for $i>1$ then $(1+a_1 t+a_2 t^2+ ... +a_n t^n) \approx (1+a_1 t)$ and 

$S=\frac{5\pi \sqrt{1-\alpha^2}}{384}(1+a_1 t)^3 S_{BH}$

\begin{figure}[ht]
\begin{center}

~~~~~~~Fig.4a~~~~~~~~~~~~~~~~~~~~~\\
\includegraphics[height=2.5in, width=3.2in]{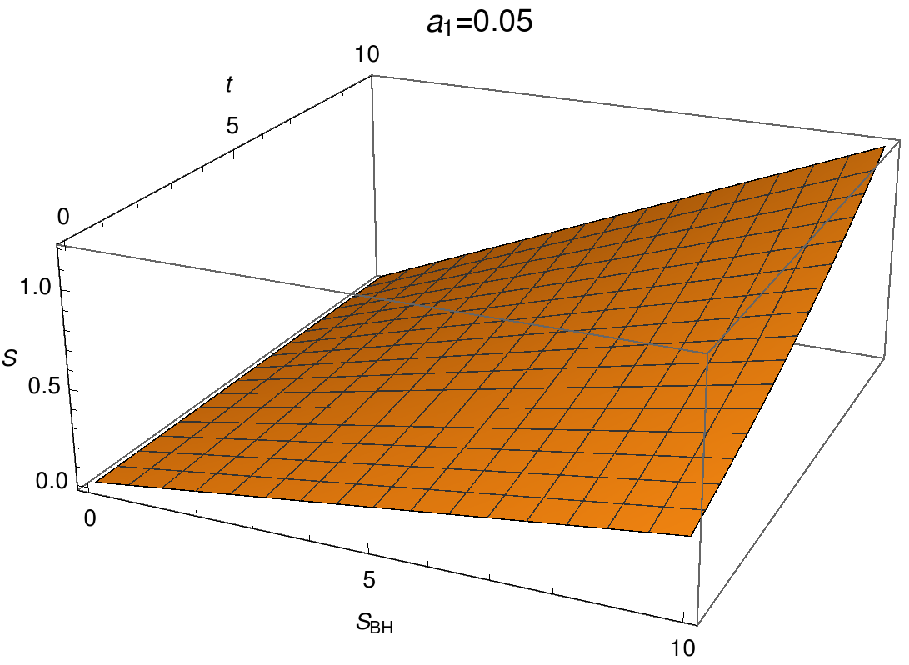}\\
\vspace{.1cm}
Relation between $S$, $S_{BH}$ and $t$ for $\alpha=0.5$

\end{center}
\end{figure}

\begin{figure}[ht]
\begin{center}

~~~~~~~Fig.4b~~~~~~~~~~~~~~~~~~~~~\\
\includegraphics[height=2.5in, width=3.2in]{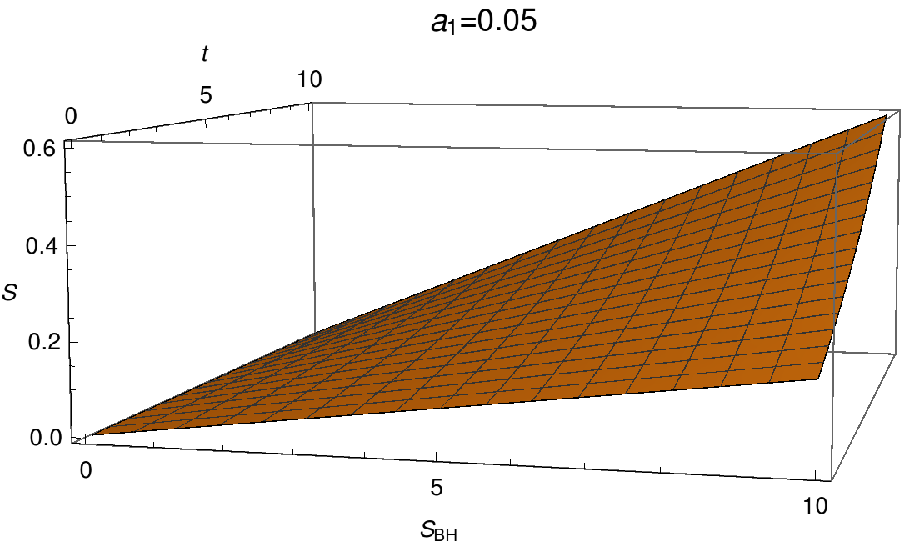}\\
\vspace{.1cm}
Relation between $S$, $S_{BH}$ and $t$ for $\alpha=0.9$

\end{center}
\end{figure}
\newpage
We have drawn the newly calculated entropy with respect to $S_{BH}$ and time for $\alpha=0.5$ and $0.9$ for $a_1=0.05$ in fig 4a and 4b respectively. We see as we increase $\alpha$, the value of entropy decreases. This means as the system is moving with a speed near to the speed of light, the variation of entropy decreases lower than the classical terminal results.

{\bf Case:II}

If $a_i=0$ for $i>2$ then  $(1+a_1 t+a_2 t^2+ ... +a_n t^n) \approx (1+a_1 t+a_2 t^2)$ and

$S=\frac{5\pi \sqrt{1-\alpha^2}}{384}(1+a_1 t+a_2 t^2)^3 S_{BH}$
 
\begin{figure}[ht]
\begin{center}

~~~~~~~Fig.5a~~~~~~~~~~~~~~~~~~~~~\\
\includegraphics[height=2.5in, width=3.2in]{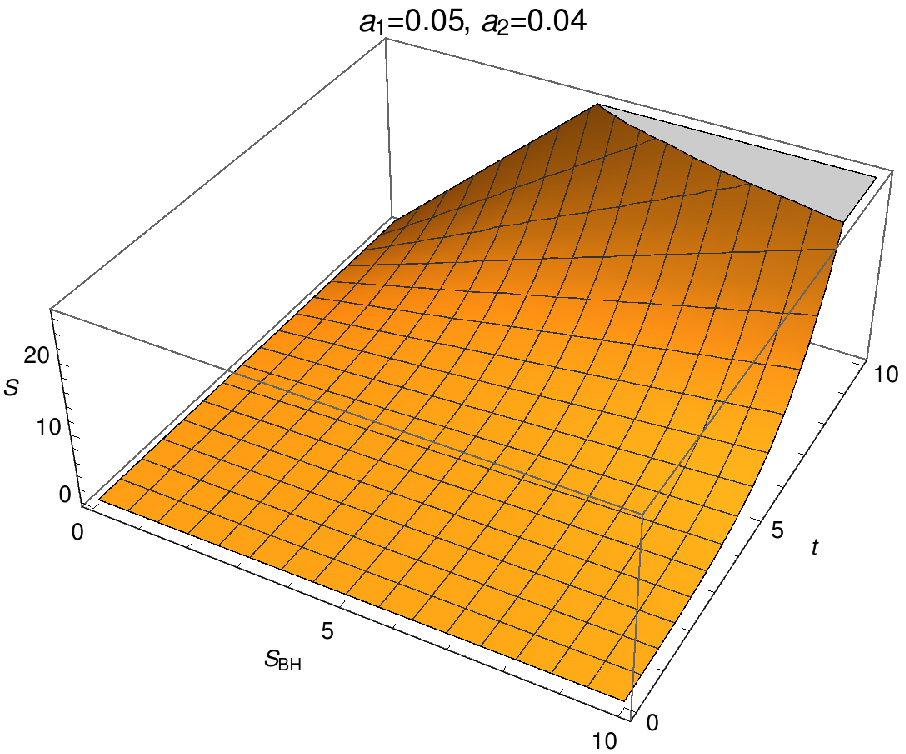}\\
\vspace{.1cm}
Relation between $S$, $S_{BH}$ and $t$ for $\alpha=0.5$

\end{center}
\end{figure}
\begin{figure}[ht]
\begin{center}

~~~~~~~Fig.5b~~~~~~~~~~~~~~~~~~~~~\\
\includegraphics[height=2.5in, width=3.2in]{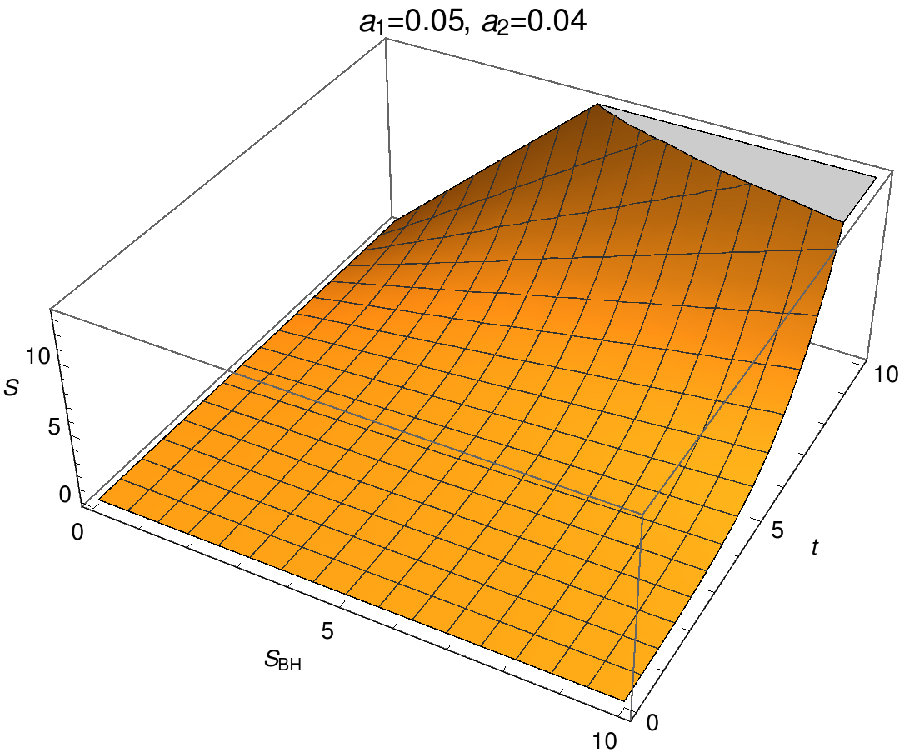}\\
\vspace{.1cm}
Relation between $S$, $S_{BH}$ and $t$ for $\alpha=0.9$

\end{center}
\end{figure}
\newpage
We have plotted the newly calculated entropy for $\alpha=0.5$ and $0.9$ for $a_1=0.05,~a_2=0.04$ and we can see in this case with time the increment of $S$ is higher than the classical case.
 
{\bf Case:III}

If $a_i=0$ for $i>2$ then  $(1+a_1 t+a_2 t^2+ ... +a_n t^n) \approx (1+a_1 t+a_2 t^2+a_3 t^3)$ and

$S=\frac{5\pi \sqrt{1-\alpha^2}}{384}(1+a_1 t+a_2 t^2+a_3 t^3)^3 S_{BH}$

\begin{figure}[ht]
\begin{center}

~~~~~~~Fig.6a~~~~~~~~~~~~~~~~~~~~~\\
\includegraphics[height=2.5in, width=3.2in]{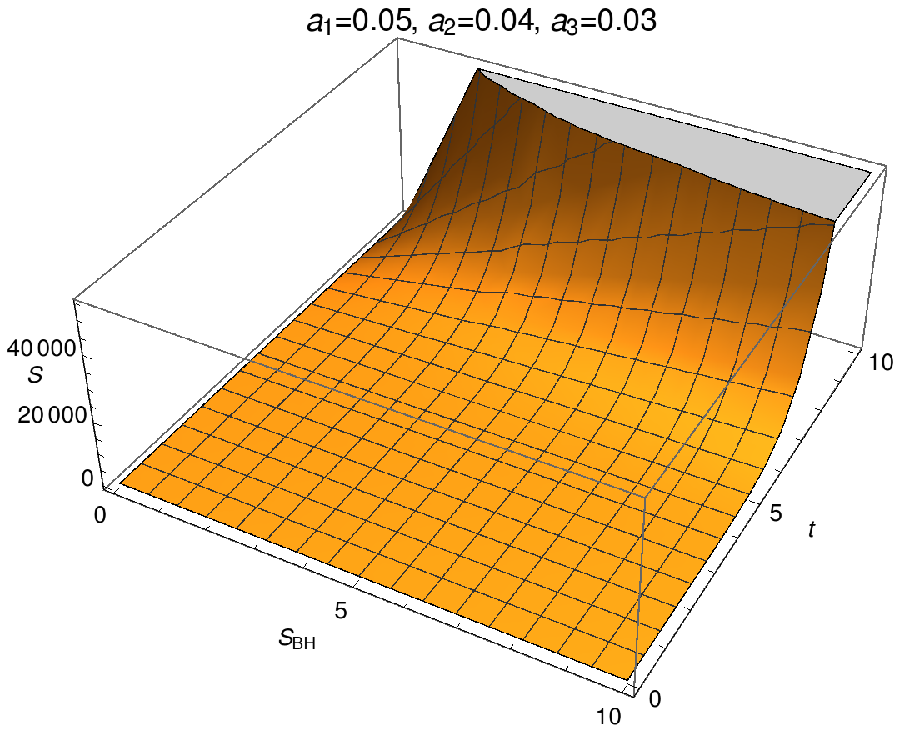}\\
\vspace{.1cm}
Relation between $S$, $S_{BH}$ and $t$ for $\alpha=0.5$

\end{center}
\end{figure}

\begin{figure}[ht]
\begin{center}

~~~~~~~Fig.6b~~~~~~~~~~~~~~~~~~~~~\\
\includegraphics[height=2.5in, width=3.2in]{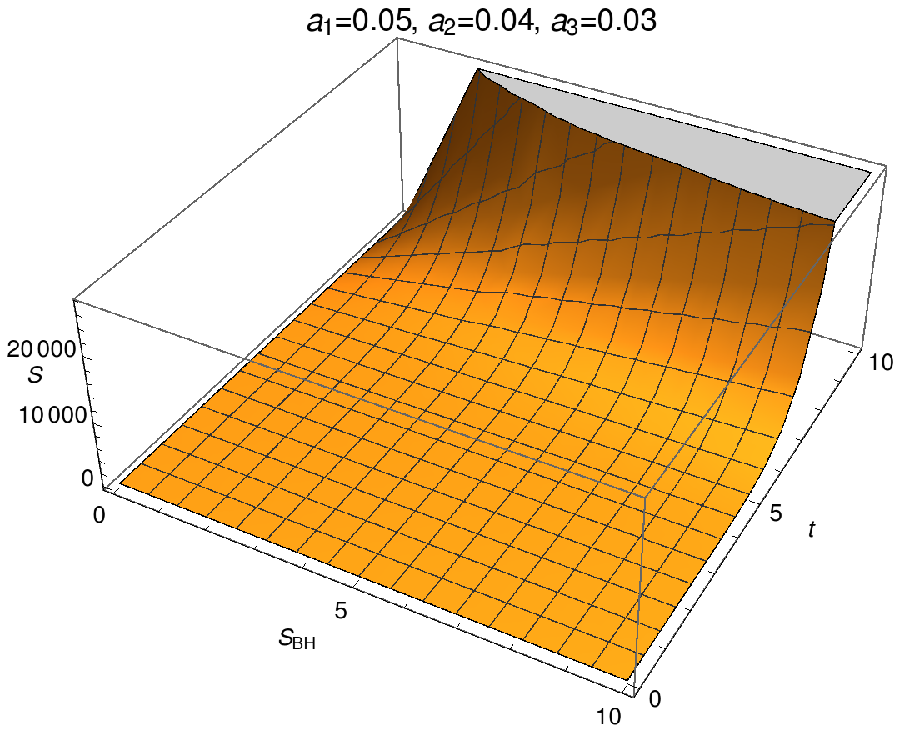}\\
\vspace{.1cm}
Relation between $S$, $S_{BH}$ and $t$ for $\alpha=0.9$

\end{center}
\end{figure}
We have plotted $S$ vs $S_{BH}$ and $t$ for $\alpha=0.5$ and $\alpha=0.9$ in the figures 6a and 6b respectively. When $a_1,~a_2$ and $a_3$ all are non zero we observe that the rate of increment of $S$ with time is higher than the previous cases.
\newpage
Fig.4a, fig.5a, fig.6a and fig.4b, fig.5b, fig.6b represents the relations between $S$, $S_{BH}$ and $t$ for $\alpha=0.5$ and $\alpha=0.9$ respectively. Also, the figures which we will get in Case:I, Case:II and CaseIII for $\alpha=0$, are identical as fig.1a, fig.2a and fig.3a respectively in section 4. The medium moves with a constant velocity $\vec{v}$ with respect to the inertial frame $K$. If the velocity $\vec{v}$ is negligible to the speed of light that case represented by taking $\alpha=0$. Also, taking $\alpha=0.5$ for those medium which moves with comparable velocity $\vec{v}$ to the speed of light. Again, if the medium moves with a velocity near to the speed of light then consider the case for $\alpha=0.9$. From the figures we can say that if we increase the velocity of the medium then the entropy decrease very fast, i.e., if we decrease the value of $\alpha$ then the entropy increase.

\section{Black Holes with Scalar Charge}
Let us consider a new type of black hole solution, namely, massive gravity black hole\cite{Bebronne1}. The ansatz for the static spherically symmetric black hole solutions can be written in the following form:
\begin{equation}
ds^2=-g(r,t)dt^2+\frac{dr^2}{g(r,t)}+r^2\left(d\theta^2+sin^2\theta d\phi^2\right)~~,
\end{equation}
where $g(r,t)=1-\frac{2M(1+a_1 t+a_2 t^2+ ... +a_n t^n)}{r}-\frac{Q}{r^{\lambda}}$ and $\lambda$ is a parameter of the model which depends on the scalar charge $Q$. The presence of the scalar charge represents a modification of the Einstein's gravitational theory. In this paper we shall focus on the case $Q>0$ and $\lambda=1$.

From the equation (\ref{Gibbs equation1}), we get, the Gibb's free energy as  
$$G_0=\frac{\pi  \left(\frac{a \beta }{4 \pi }+Q\right)^3}{576 \beta ^2}\Bigg[60 \sin ^{-1}\left(\sqrt{\frac{a \beta }{4 \pi }+Q}\right)-60 \sin ^{-1}\left(\frac{Q}{\sqrt{\frac{a \beta }{4 \pi }+Q}}\right)-45
   \sin \left\{2 \sin ^{-1}\left(\frac{Q}{\sqrt{\frac{a \beta }{4 \pi }+Q}}\right)\right\}$$
$$-9 \sin \left\{4 \sin ^{-1}\left(\frac{Q}{\sqrt{\frac{a \beta }{4 \pi }+Q}}\right)\right\}-\sin \left\{6 \sin
   ^{-1}\left(\frac{Q}{\sqrt{\frac{a \beta }{4 \pi }+Q}}\right)\right\}+45 \sin \left\{2 \sin ^{-1}\left(\sqrt{\frac{a \beta }{4 \pi
   }+Q}\right)\right\}$$  
\begin{equation}
+9 \sin \left\{4 \sin ^{-1}\left(\sqrt{\frac{a \beta }{4 \pi }+Q}\right)\right\}+\sin \left\{6 \sin ^{-1}\left(\sqrt{\frac{a \beta }{4
   \pi }+Q}\right)\right\}\Bigg]~~~,
\end{equation}  
where $a=(1+a_1 t+a_2 t^2+ ... +a_n t^n)$. 
  
Hence, the entropy is
$$S_0=\frac{1}{576} \pi  \left(\frac{a \beta }{4 \pi }+Q\right)^3\Bigg[\frac{15 a Q}{2 \pi  \left(\frac{a \beta }{4 \pi }+Q\right)^{3/2}
   \sqrt{1-\frac{Q^2}{\frac{a \beta }{4 \pi }+Q}}}+\frac{45 a Q \cos \left\{2
   \sin ^{-1}\left(\frac{Q}{\sqrt{\frac{a \beta }{4 \pi
   }+Q}}\right)\right\}}{4 \pi  \left(\frac{a \beta }{4 \pi }+Q\right)^{3/2}
   \sqrt{1-\frac{Q^2}{\frac{a \beta }{4 \pi }+Q}}}+\frac{15 a}{2 \pi 
   \sqrt{1-\frac{a \beta }{4 \pi }-Q} \sqrt{\frac{a \beta }{4 \pi }+Q}}$$
$$+\frac{9 a Q \cos \left\{4 \sin ^{-1}\left(\frac{Q}{\sqrt{\frac{a \beta }{4
   \pi }+Q}}\right)\right\}}{2 \pi  \left(\frac{a \beta }{4 \pi
   }+Q\right)^{3/2} \sqrt{1-\frac{Q^2}{\frac{a \beta }{4 \pi }+Q}}}+\frac{3
   a Q \cos \left\{6 \sin ^{-1}\left(\frac{Q}{\sqrt{\frac{a \beta }{4 \pi
   }+Q}}\right)\right\}}{4 \pi  \left(\frac{a \beta }{4 \pi }+Q\right)^{3/2}
   \sqrt{1-\frac{Q^2}{\frac{a \beta }{4 \pi }+Q}}}+\frac{45 a \cos \left\{2
   \sin ^{-1}\left(\sqrt{\frac{a \beta }{4 \pi }+Q}\right)\right\}}{4 \pi 
   \sqrt{1-\frac{a \beta }{4 \pi }-Q} \sqrt{\frac{a \beta }{4 \pi }+Q}}$$
$$+\frac{9 a \cos \left\{4 \sin ^{-1}\left(\sqrt{\frac{a \beta }{4 \pi
   }+Q}\right)\right\}}{2 \pi  \sqrt{1-\frac{a \beta }{4 \pi }-Q}
   \sqrt{\frac{a \beta }{4 \pi }+Q}}+\frac{3 a \cos \left\{6 \sin
   ^{-1}\left(\sqrt{\frac{a \beta }{4 \pi }+Q}\right)\right\}}{4 \pi 
   \sqrt{1-\frac{a \beta }{4 \pi }-Q} \sqrt{\frac{a \beta }{4 \pi }+Q}}\Bigg]$$ 
$$+\frac{1}{768} a \left(\frac{a \beta }{4 \pi }+Q\right)^2\Bigg[60
   \sin ^{-1}\left(\sqrt{\frac{a \beta }{4 \pi }+Q}\right)-60 \sin ^{-1}\left(\frac{Q}{\sqrt{\frac{a \beta }{4 \pi }+Q}}\right)-45 \sin \left\{2
   \sin ^{-1}\left(\frac{Q}{\sqrt{\frac{a \beta }{4 \pi
   }+Q}}\right)\right\}$$
$$-9 \sin \left\{4 \sin ^{-1}\left(\frac{Q}{\sqrt{\frac{a
   \beta }{4 \pi }+Q}}\right)\right\}+45 \sin \left\{2 \sin ^{-1}\left(\sqrt{\frac{a \beta }{4 \pi
   }+Q}\right)\right\}-\sin \left\{6 \sin ^{-1}\left(\frac{Q}{\sqrt{\frac{a
   \beta }{4 \pi }+Q}}\right)\right\}$$
$$+9 \sin \left\{4 \sin ^{-1}\left(\sqrt{\frac{a \beta }{4 \pi
   }+Q}\right)\right\}+\sin \left\{6 \sin ^{-1}\left(\sqrt{\frac{a \beta }{4
   \pi }+Q}\right)\right\}\Bigg]$$ 
$$-\frac{\pi  \left(\frac{a \beta }{4 \pi }+Q\right)^3}{288 \beta }\Bigg[60
   \sin ^{-1}\left(\sqrt{\frac{a \beta }{4 \pi }+Q}\right)-60 \sin ^{-1}\left(\frac{Q}{\sqrt{\frac{a \beta }{4 \pi }+Q}}\right)-45 \sin \left\{2
   \sin ^{-1}\left(\frac{Q}{\sqrt{\frac{a \beta }{4 \pi }+Q}}\right)\right\}$$ 
$$-9 \sin \left\{4 \sin ^{-1}\left(\frac{Q}{\sqrt{\frac{a \beta }{4 \pi
   }+Q}}\right)\right\}-\sin \left\{6 \sin ^{-1}\left(\frac{Q}{\sqrt{\frac{a
   \beta }{4 \pi }+Q}}\right)\right\}+45 \sin \left\{2 \sin
   ^{-1}\left(\sqrt{\frac{a \beta }{4 \pi }+Q}\right)\right\}$$ 
\begin{equation}
+9 \sin \left\{4 \sin ^{-1}\left(\sqrt{\frac{a \beta }{4 \pi
   }+Q}\right)\right\}+\sin \left\{6 \sin ^{-1}\left(\sqrt{\frac{a \beta }{4
   \pi }+Q}\right)\right\}\Bigg]
\end{equation}
The relation between $S_0$, $S_{BH}$ and $t$   looks like             
$$S_0=\frac{1}{576} \pi  \left(a \sqrt{S_{BH}}+Q\right)^3\Bigg[\frac{15 a Q}{2 \pi  \sqrt{1-\frac{Q^2}{a \sqrt{S_{BH}}+Q}} \left(a \sqrt{S_{BH}}+Q\right)^{3/2}}+\frac{45 a Q \cos
   \left\{2 \sin ^{-1}\left(\frac{Q}{\sqrt{a \sqrt{S_{BH}}+Q}}\right)\right\}}{4 \pi  \sqrt{1-\frac{Q^2}{a
   \sqrt{S_{BH}}+Q}} \left(a \sqrt{S_{BH}}+Q\right){}^{3/2}}$$
$$+\frac{15 a}{2 \pi  \sqrt{1-a\sqrt{S_{BH}}-Q} \sqrt{a \sqrt{S_{BH}}+Q}}+\frac{9 a Q \cos \left\{4 \sin ^{-1}\left(\frac{Q}{\sqrt{a \sqrt{S_{BH}}+Q}}\right)\right\}}{2 \pi  \sqrt{1-\frac{Q^2}{a
   \sqrt{S_{BH}}+Q}} \left(a \sqrt{S_{BH}}+Q\right)^{3/2}}+\frac{3 a Q \cos \left\{6 \sin ^{-1}\left(\frac{Q}{\sqrt{a
   \sqrt{S_{BH}}+Q}}\right)\right\}}{4 \pi  \sqrt{1-\frac{Q^2}{a \sqrt{S_{BH}}+Q}} \left(a
   \sqrt{S_{BH}}+Q\right)^{3/2}}$$
$$+\frac{45 a \cos \left\{2 \sin ^{-1}\left(\sqrt{a \sqrt{S_{BH}}+Q}\right)\right\}}{4 \pi  \sqrt{1-a\sqrt{S_{BH}}-Q} \sqrt{a \sqrt{S_{BH}}+Q}}+\frac{9 a \cos \left\{4 \sin ^{-1}\left(\sqrt{a
   \sqrt{S_{BH}}+Q}\right)\right\}}{2 \pi  \sqrt{1-a\sqrt{S_{BH}}-Q} \sqrt{a
   \sqrt{S_{BH}}+Q}}$$ 
$$+\frac{3 a \cos \left\{6 \sin ^{-1}\left(\sqrt{a \sqrt{S_{BH}}+Q}\right)\right\}}{4 \pi  \sqrt{1-a\sqrt{S_{BH}}-Q} \sqrt{a \sqrt{S_{BH}}+Q}}\Bigg]+\frac{1}{768} a \left(a \sqrt{S_{BH}}+Q\right)^2\Bigg[60 \sin ^{-1}\left(\sqrt{a \sqrt{S_{BH}}+Q}\right)$$
$$-60 \sin ^{-1}\left(\frac{Q}{\sqrt{a \sqrt{S_{BH}}+Q}}\right)-45 \sin \left\{2 \sin ^{-1}\left(\frac{Q}{\sqrt{a
   \sqrt{S_{BH}}+Q}}\right)\right\}-9 \sin \left\{4 \sin ^{-1}\left(\frac{Q}{\sqrt{a \sqrt{S_{BH}}+Q}}\right)\right\}$$    
$$-\sin \left\{6 \sin ^{-1}\left(\frac{Q}{\sqrt{a \sqrt{S_{BH}}+Q}}\right)\right\}+45 \sin \left\{2 \sin ^{-1}\left(\sqrt{a
   \sqrt{S_{BH}}+Q}\right)\right\}+9 \sin \left\{4 \sin ^{-1}\left(\sqrt{a \sqrt{S_{BH}}+Q}\right)\right\}$$ 
$$+\sin \left\{6 \sin ^{-1}\left(\sqrt{a \sqrt{S_{BH}}+Q}\right)\right\}\Bigg]-\frac{1}{288} \pi  \left(a \sqrt{S_{BH}}+Q\right)^3\Bigg[60 \sin ^{-1}\left(\sqrt{a \sqrt{S_{BH}}+Q}\right)$$ 
$$-60 \sin ^{-1}\left(\frac{Q}{\sqrt{a \sqrt{S_{BH}}+Q}}\right)-45 \sin \left\{2 \sin ^{-1}\left(\frac{Q}{\sqrt{a
   \sqrt{S_{BH}}+Q}}\right)\right\}-9 \sin \left\{4 \sin ^{-1}\left(\frac{Q}{\sqrt{a \sqrt{S_{BH}}+Q}}\right)\right\}$$ 
$$-\sin \left\{6 \sin ^{-1}\left(\frac{Q}{\sqrt{a \sqrt{S_{BH}}+Q}}\right)\right\}+45 \sin \left\{2 \sin ^{-1}\left(\sqrt{a
   \sqrt{S_{BH}}+Q}\right)\right\}+9 \sin \left\{4 \sin ^{-1}\left(\sqrt{a \sqrt{S_{BH}}+Q}\right)\right\}$$ 
\begin{equation}
+\sin \left\{6 \sin ^{-1}\left(\sqrt{a \sqrt{S_{BH}}+Q}\right)\right\}\Bigg]
\end{equation}         

{\bf Case:I}

If $a_i=0$ for $i>1$ then $(1+a_1 t+a_2 t^2+ ... +a_n t^n) \approx (1+a_1 t)$                     

\begin{figure}[ht]
\begin{center}

~~~~~~~Fig.7a~~~~~~~~~~~~~~~~~~~~~\\
\includegraphics[height=2.5in, width=3.2in]{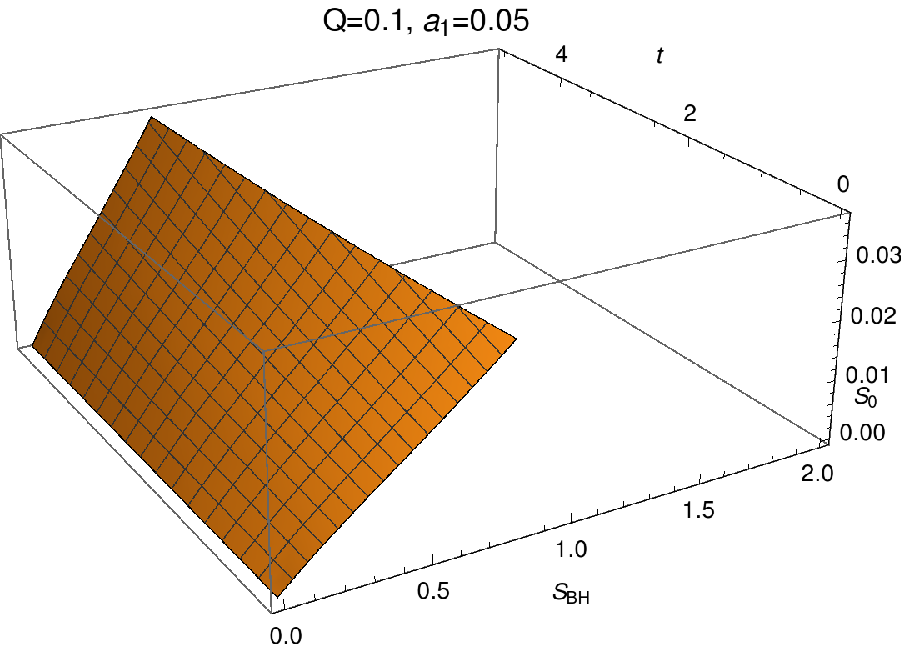}\\
\vspace{.1cm}
Relation between $S_0$, $S_{BH}$ and $t$

\end{center}
\end{figure}
{\bf Case:II}

If $a_i=0$ for $i>2$ then $(1+a_1 t+a_2 t^2+ ... +a_n t^n) \approx (1+a_1 t+a_2 t^2)$                     
\begin{figure}[ht]
\begin{center}

~~~~~~~Fig.7b~~~~~~~~~~~~~~~~~~~~~\\
\includegraphics[height=2.5in, width=3.2in]{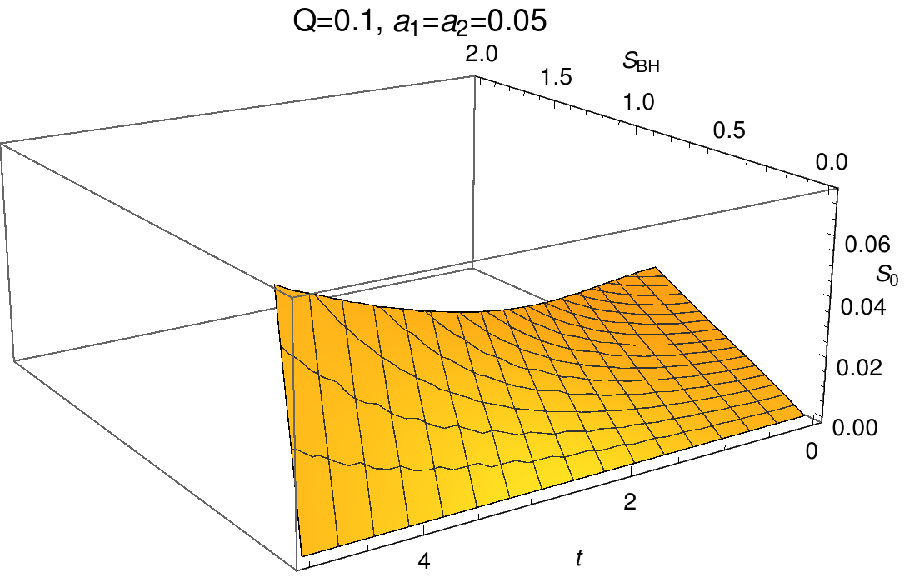}\\
\vspace{.1cm}
Relation between $S_0$, $S_{BH}$ and $t$

\end{center}
\end{figure}
\newpage
{\bf Case:III}

If $a_i=0$ for $i>3$ then $(1+a_1 t+a_2 t^2+ ... +a_n t^n) \approx (1+a_1 t+a_2 t^2+a_3 t^3)$                     
\begin{figure}[ht]
\begin{center}

~~~~~~~Fig.7c~~~~~~~~~~~~~~~~~~~~~\\
\includegraphics[height=2.5in, width=3.2in]{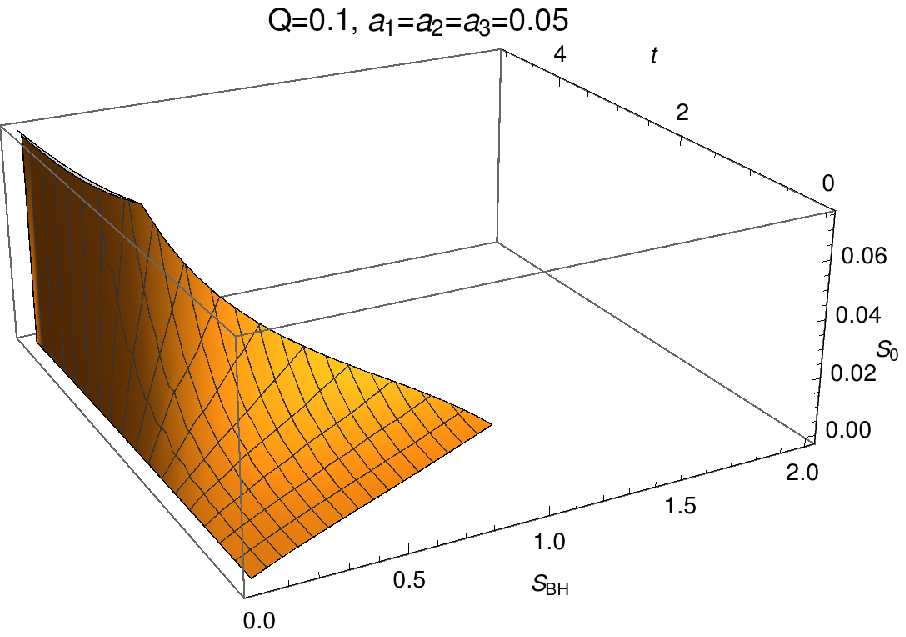}\\
\vspace{.1cm}
Relation between $S_0$, $S_{BH}$ and $t$

\end{center}
\end{figure}

Relation between $S_0$, $S_{BH}$ and $t$ for Case-I, Case-II and Case-III are shown in the figures $7a-7c$.

\section{Conclusion}
%%%%%%%%%%%%%%%%%%%%%%%%%%%%%%%%%%%%%%%%%%%%%%%%%%%%%%%%%%%%
A black hole is usually formed from the collapse of a quantity of matter or radiation, both of which carry entropy. However, the black hole's interior and contents are veiled to an exterior observer. Thus a thermodynamic description of the collapse from that observer's view point cannot be based on the entropy of that matter or radiation because these are unobservable. A stationary black hole is parametrized by just a few numbers (Ruffini and Wheeler 1971): its mass, electric charge and angular momentum (and magnetic monopole charge, except its actual existence in nature has not been demonstrated yet). For any specific choice of these parameters one can imagine many scenarios for the black hole's formation. Thus there are many possible internal states corresponding to that black hole. In thermodynamics one meets a similar situation: many internal microstates of a system are all compatible with the one observed (macro) state. Thermodynamic entropy quantifies the said multiplicity. Thus by analogy one needs to associate entropy with a black hole . 

In this paper, at first we calculated the entropy for the Schwarzchild black hole contained by the CR volume for massless modes. The approach are considered as statistical one. We first introduced the integrand of the expression of interior volume as an effective metric. Then easily identified the energy of the modes. To handle the situation, one can use the method of constraint analysis, since the canonical Hamiltonian vanishes. The exact calculation of the Hamiltonian is itself new. By using the Gibb's free energy we can calculate the entropy. The effect is very interesting. As the mass of the black hole increasing as time increases, the average density of the universe is very small. So, the increment of the mass of the black hole is very slow. So we consider the mass of the black hole is $M(1+a_1 t+a_2 t^2+ ... +a_n t^n)$ after a time $t$. Then we assumed $r$ varies from $2M(1+a_1 t+a_2 t^2+ ... +a_n t^n)$ to zero. Finally we found that the entropy is monotonically increasing as time increases but the increment of entropy depends on time which is very small. Also we can see that the entropy on the horizon is greater than the entropy inside the black hole i.e. $S_{BH} > S$. The entropy inside the black hole is proportional to the time($t$) i.e. if we increase the time then the entropy($S$) inside the black hole increase. Hope this paper will give a imagination knowledge about entropy. 

%%%%%%%%%%%%%%%%%%%%%%%%%%%%%%%%%%%%%%%%%%%%%%%%%%%

\vspace{1 in}
{\bf Acknowledgement:}
This research is Supported by the project grant of Goverment of West Bengal, Department of Higher Education, Science and Technology and Biotecnology (File no:- $ ST/P/S \& T/16G-19/2017$). SD thanks West Bengal State Govt. for awarding Non-NET Fellowship. RB thanks Inter University Center for Astronomy and Astrophysics(IUCAA), Pune, India for Visiting Associateship.
%%%%%%%%%%%%%%%%%%%%%%%%%%%%%%%%%%%%%%%%%%%%%%%%%

\end{document}